\documentclass[twocolumn,showpacs,preprintnumbers,superscriptaddress,amsmath,amssymb]{revtex4}

\usepackage{graphicx}
\usepackage{pgf}

\usepackage{epsfig,latexsym,amssymb,color,bm}

\usepackage{booktabs,longtable,mathrsfs}
\usepackage{multirow}

\date{\today}

\newcommand{\SKM}{SkM$^\ast$}
\newcommand{\SLY}{SLy4$_{\rm L}$}

\begin{document}

\title{Self-consistent tilted-axis-cranking study of
triaxial strongly deformed bands in $^{158}$Er at ultrahigh spin}

\author{Yue Shi}
\affiliation{State Key Laboratory of
Nuclear Physics and Technology, School of Physics, Peking University, Beijing 100871, China}
\affiliation{Department of Physics and Astronomy, University of
Tennessee, Knoxville, Tennessee 37996, USA}
\affiliation{Physics Division, Oak Ridge National Laboratory, Post
Office Box 2008, Oak Ridge, Tennessee 37831, USA}
\affiliation{Department of Physics, PO Box 35 (YFL), FI-40014
University of Jyv{\"a}skyl{\"a}, Finland}

\author{J. Dobaczewski}
\affiliation{Institute of Theoretical Physics, Faculty of Physics, University of Warsaw, ul. Ho{\.z}a 69, PL-00681 Warsaw, Poland}
\affiliation{Department of Physics, PO Box 35 (YFL), FI-40014
University of Jyv{\"a}skyl{\"a}, Finland}

\author{S. Frauendorf}
\affiliation{Department of Physics, University of Notre Dame, Notre
Dame, Indiana 46556, USA}

\author{W. Nazarewicz}
\affiliation{Department of Physics and Astronomy, University of
Tennessee, Knoxville, Tennessee 37996, USA}
\affiliation{Physics Division, Oak Ridge National Laboratory, Post
Office Box 2008, Oak Ridge, Tennessee 37831, USA}
\affiliation{Institute of Theoretical Physics, Faculty of Physics, University of Warsaw, ul.
Ho{\.z}a 69, PL-00681 Warsaw, Poland}

\author{J.C. Pei}
\affiliation{Joint Institute for Heavy-Ion Research, Oak Ridge,
Tennessee 37831, USA}
\affiliation{Department of Physics and Astronomy, University of
Tennessee, Knoxville, Tennessee 37996, USA}
\affiliation{Physics Division, Oak Ridge National Laboratory, Post
Office Box 2008, Oak Ridge, Tennessee 37831, USA}

\author{F.R. Xu}
\affiliation{State Key Laboratory of
Nuclear Physics and Technology, School of Physics, Peking University, Beijing 100871,
China}

\author{N. Nikolov}
\affiliation{Department of Physics and Astronomy, University of
Tennessee, Knoxville, Tennessee 37996, USA}

\begin{abstract}

Stimulated by  recent experimental discoveries,  triaxial strongly deformed (TSD) states in $^{158}$Er at ultrahigh spins have been  studied by means of the Skyrme-Hartree-Fock model and the tilted-axis-cranking method. Restricting the rotational axis to one of the principal axes -- as done in   previous cranking calculations -- two well-defined TSD minima in the total Routhian surface are found for a given configuration: one with  positive and another with negative triaxial deformation $\gamma$. By allowing the rotational axis to change direction, the higher-energy minimum is shown to be a saddle point. This resolves the long-standing question of the physical interpretation of the two triaxial minima  at a very similar quadrupole shape obtained in the principal axis cranking approach.
Several TSD configurations  have been  predicted, including a highly deformed band  expected to cross lesser elongated TSD bands  at the highest spins. Its transitional quadrupole moment $Q_t \approx 10.5$\,eb is close to the measured value of $\sim$11\,eb; hence, it is a candidate for the structure observed in experiment.
\end{abstract}


\maketitle

While the majority of nuclei have axially symmetric
shapes,  evidence for triaxial nuclear deformations  has been elusive. The clearest signatures come from the gamma-ray spectroscopy of rotating nuclei.
The deformation of a quantum object, such as molecule or atomic
nucleus, enables the system to specify an orientation. The quantized motion
of this degree of freedom generates the sequences of rotational levels - the rotational
bands~\cite{bohr75,frau01}. If the system is triaxial, the associated rotational bands
show specific features that allow for distinguishing it from an axial one.
In the case of nuclei, the appearance of the wobbling~\cite{bohr75,odeg01,jen02} and spin-chirality~\cite{frau97,frau01}  rotational modes are
experimental signatures of triaxiality.

Triaxial shapes are expected to
appear more frequently at high spin because of
the tendency of aligned high-$j$ quasiparticles to drive rotating nuclei towards triaxiality due to their spatial density distributions \cite{framay83,mate07}. In addition, pairing correlations -- which generally favor more symmetric shapes -- are quenched at high spins and enhance  the high-$j$ alignment effect \cite{mats02}. Consequently, with increasing spin, nuclei are predicted
to go through nonaxial shapes before they eventually fission (see, e.g., \cite{werner92}).

Recent experiments~\cite{paul07,patt07,agui08,tea08,oll09,oll11,wang11} have reached ultrahigh spins of about $65\hbar$ in nuclei around $^{158}$Er. It has been observed that with increasing angular momentum,  the rotational bands terminate and nuclei assume weakly deformed oblate shapes, as evidenced by the irregular level spacings. At ultrahigh spins they return to collective rotation characterized  by regular rotational bands, consistent with the early prediction \cite{dude85}. Cranked Nilsson-Strutinsky (CNS) calculations suggest that the observed bands in $^{158}$Er are based on one of the three triaxial strongly deformed (TSD) minima  in the potential-energy surface  (see Ref.~\cite{wang11} and references cited therein). For the lowest minimum, TSD1,  the calculated value of the transitional quadrupole moment $Q_t \approx 7.5$\,eb considerably underestimates the observed value of $\sim 11$\,eb~\cite{wang11}. This has led to the suggestion that the observed band in $^{158}$Er may be associated with either the minimum TSD2, which has a similar quadrupole deformation parameter $\varepsilon_2$  as TSD1  but  negative $\gamma$, or with the band TSD3, which has a larger  triaxial deformation~\cite{wang11}.

Most of the existing high-spin calculations in the mass-160 region assume that the axis of rotation coincides with one of the principal axes of the triaxial potential, which is commonly referred to as principal-axis cranking (PAC). The CNS calculations use the microscopic-macroscopic method, which combines a shell correction derived from a phenomenological potential with the deformation energy of a rotating liquid drop~\cite{beng85,afan99,dude85}. It is common to choose the $x$-axis as the rotational axis and let the triaxiality parameter cover the range $-120^\circ\leq\gamma\leq60^\circ$. In the Lund convention, which we adopt in this Letter, the three sectors $[-120^\circ,-60^\circ]$,   $[-60^\circ,0^\circ]$, and $[0^\circ,60^\circ]$ represent the same triaxial shapes but represent rotation about the long, medium and short axis, respectively. The TSD1 and TSD2 minima in the CNS calculations~\cite{wang11} correspond to similar values of $\varepsilon_2$ and $|\gamma|$ which means that their shapes are nearly the same. The opposite sign of  $\gamma$ means that TSD1 rotates about the short axis and TSD2 about the medium axis. This raises the question of their physical interpretation, e.g., whether the higher of the two minima obtained in PAC is stable with respect to a reorientation of the rotational axis.

In this Letter we address this question by means of the tilted axis cranking (TAC) method \cite{frau93,kerm81},
which considers the general orientation of  the axis of rotation with respect to
the principal axes of the nuclear quadrupole moment. We  investigate the structure of
ultrahigh-spin TSD minima in $^{158}$Er by using two approaches: the shell-correction
tilted axis cranking method (SCTAC)   \cite{frau93}, which is based on
the phenomenological Nilsson potential, and, for the first time,  the three-dimensional
self-consistent Skyrme-Hartree-Fock version of tilted axis cranking (SHFTAC)
developed in Ref.~\cite{olbr04}.
Employing a fully self-consistent rotating mean field -- including the full rotational response due to cranking --  is expected to improve the reliability
of calculations in the realm of ultrahigh-spin states.

SHFTAC is based  on the
symmetry-unrestricted solver {\sc hfodd} (v2.49s)~\cite{schu11}, which has been successfully applied
to the description of chiral bands in $^{132}$La \cite{olbr04}.
In the particle-hole channel, we use the Skyrme energy density functionals
{\SKM}~\cite{bart82} and SLy4~\cite{chab98}, the
latter of which has been supplemented with Landau parameters ({\SLY})~\cite{bend02,zdu05}.  The total energy of the system ${\cal E}$ is obtained by integrating the total energy density over spatial coordinates. We have used 1,000
 deformed harmonic oscillator basis states with $\hbar
\omega_{\bot} = 10.080$\,MeV and $\hbar \omega_{\parallel}=7.418$\,MeV.
At ultrahigh spins, pairing
is negligible; hence, it has been ignored in  SCTAC and SHFTAC.
As discussed in  earlier  SHFPAC (a PAC limit of SHFTAC) calculations \cite{mate07}, quadrupole polarization at high spin -- both axial and triaxial -- is very well described by unpaired theory.
\begin{figure}[htb] \centering
\epsfxsize=2in
\epsffile{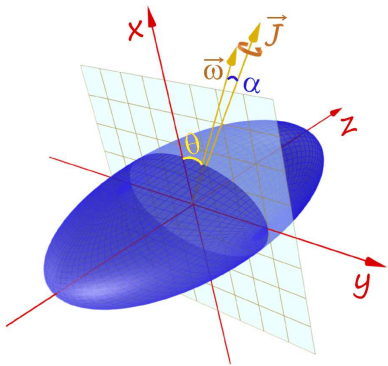}
        \caption{(Color online) Schematic picture of a TSD shape.
	The angles $\theta$ (between the $x$-axis and the rotational axis) and $\alpha$ (between $\bm{\omega}$ and $\bm{J}$) are defined
        in the $x-y$ plane. The short, medium, and long axes are
        denoted by $x$, $y$, and $z$, respectively; that is, the plotted shape corresponds to $\gamma>0$.
} \label{shape2}
\end{figure}
The SCTAC calculations give a strong increase of the Routhian
${\cal E}^{\bm{\omega}}\equiv {\cal E}-\bm{\omega}\cdot \bm{J}$ when the
rotational axis is tilted toward the $z$-axis. For this reason we restrict the numerically
extensive SHFTAC calculation to the $x-y$ plane spanned by the short and medium axes. The tilt angle $\theta$ of the rotational axis  is measured with respect to the short axis, as illustrated in Fig.~\ref{shape2}.
\begin{figure}[htb] \centering
\includegraphics[width=0.4\textwidth]{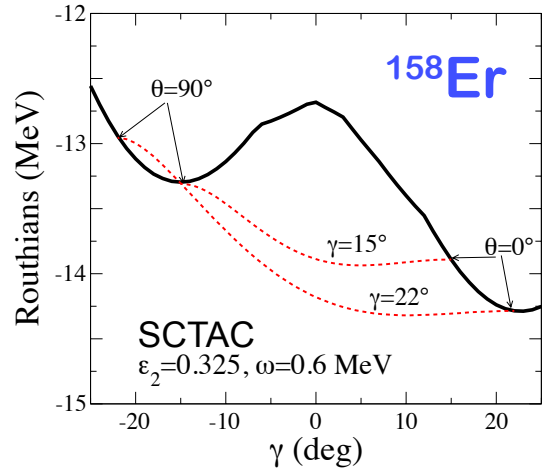}
\caption{(Color online) Lowest Routhians in $^{158}$Er calculated by means of SCTAC (Nilsson potential)  at fixed $\varepsilon_2$ and $\omega$ as a function of $\gamma$ for rotation about the $x$-axis (solid line) compared to those calculated at $\gamma=15^\circ$ and  $22^\circ$ as functions of $\theta$ (dashed lines). In the latter case, the Routhians are drawn by uniformly scaling the range of $0^\circ\leq\theta\leq90^\circ$ into the corresponding ranges of $\gamma$.
} \label{shape}
\end{figure}

Figure~\ref{shape} shows the results of the SCTAC calculations.
If the axis of rotation agrees with one of the principal axes (PAC),
SCTAC coincides essentially with the CNS of
Ref.~\cite{beng85}. (In SCTAC the Strutinsky renormalization is only
carried out for the non-rotational part of the Routhian, whereas in
CNS  the rotational energy is also renormalized.)
The equilibrium
deformation parameters in SCTAC for rotation about the short axis are $\varepsilon_2=0.325$,
$\gamma=22^{\circ}$ (lower minimum), and for rotation about the medium axis
$\varepsilon_2=0.31$, $\gamma=-15^{\circ}$ (higher minimum), which are close to the ones found by
means of the CNS for the TSD1 and TSD2 minima, respectively \cite{wang11}.

It is clearly seen in Fig.~\ref{shape}
 that the PAC minimum
at $\gamma=-15^{\circ}$ becomes  a saddle if  the
rotational axis is allowed to tilt. The dashed lines show how the
energies change in a smooth way when tilting the rotational axis from
short ($\theta=0^{\circ}$) to medium ($\theta=90^{\circ}$)
while keeping $\varepsilon_2$ and $\gamma$ constant. {\it Thus one cannot associate  TSD2
with the band observed in $^{158}$Er.} In addition, the transition quadrupole moment for the stable
minimum TSD1 is $\sim$8\,eb, which is too small as compared with the experimental value
of $\sim$11\,eb (see also the discussion in \cite{wang11}).

In the case of SCTAC, the tilt angle of the rotational axis  is
defined relative to one of the principal axes of the deformed
potential in a straightforward way. In SHFTAC calculations, it must be
introduced by means of a constraint on the orientation of $\bm J$
along with the constraints ${\rm Im}({Q}_{22})  =
{Q}_{2\pm1}  = 0$ on the orientation of the principal axis of the
nucleus defined in terms of the total (mass) quadrupole moment $Q_{2\mu}$. The  conditions on the corresponding Lagrange multipliers have been
derived by Kerman and Onishi (KO) \cite{kerm81}.
We use the 2D counterpart of relation (3.6) of
Ref.~\cite{kerm81},  which states  that $\bm{\omega}$ and $\bm{J}$
are not parallel ($\alpha \ne 0^{\circ}$) if the Routhian is not at a stationary point (e.g., PAC).
By using the Augmented Lagrangian Method of the HFODD code~\cite{schu11},
we have checked that the KO conditions are obeyed to a high precision for all angles $\theta$.  The resulting angles $\alpha$ do
not exceed  0.1--$0.2^\circ$, depending on configuration.

\begin{table}[ht]
\caption[T1]{\label{tab1}The SHF configurations in $^{158}$Er studied in this Letter.
Each configuration is described by the number of states
occupied in the four parity-signature ($\pi,r$) blocks, in the convention defined
in Ref.~\protect\cite{doba97}, and also by the transition quadrupole moment $Q_t$.
The $Q_t$ values are from the SHF-{\SKM} calculations at $\omega=0.6$\,MeV and $\theta=0^{\circ}$.
}
\begin{ruledtabular}
\begin{tabular}{ccccc}
\multicolumn{2}{c}{configuration} & $\pi$ & $r$ &\multicolumn{1}{c}{$Q_t$ (eb)}  \\
\hline
A: & $\nu[23, 23, 22, 22] \otimes \pi[17, 18, 16, 17]$ & $-$ & $-1$ & 7.68 \\
B: & $\nu[23, 23, 22, 22] \otimes \pi[17, 17, 17, 17]$ & $+$ & $+1$ & 7.71 \\
C: & $\nu[23, 24, 21, 22] \otimes \pi[17, 18, 16, 17]$ & $+$ & $+1$ & 7.36 \\
D: & $\nu[23, 23, 22, 22] \otimes \pi[17, 17, 17, 17]$ & $+$ & $+1$ & 10.72 \\
\end{tabular}
\end{ruledtabular}
\end{table}

\begin{figure} \centering
        \includegraphics[width=0.45\textwidth]{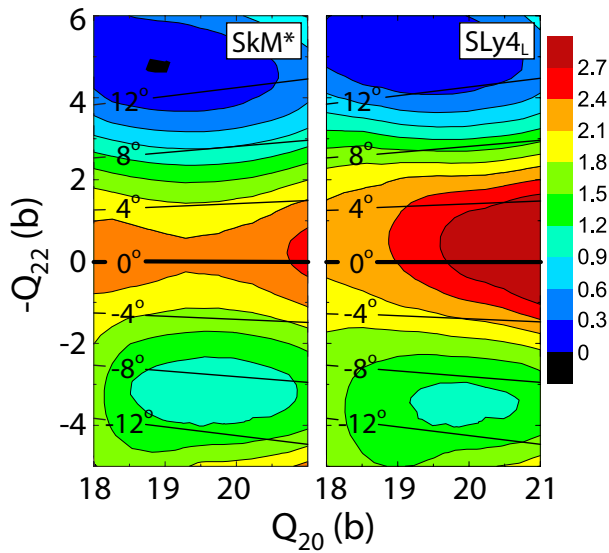}
        \caption{(Color online) Total Routhian surfaces in the ($Q_{20},Q_{22}$) plane for  the A-configuration of Table~\protect\ref{tab1}
        at $\omega=0.7$\,MeV and $\theta=0$ (PAC) obtained with
        SkM* (left) and SLy4$_{\rm L}$ (right) functionals. The corresponding values of $\gamma$ are marked.
}
        \label{PES}
\end{figure}
In this Letter, we studied four different SHF configurations
listed in Table~\ref{tab1}. They all are expected to appear near yrast around
$Q_{20}$=19\,eb.
First, in Fig.~\ref{PES} we show the SHFPAC Routhian surfaces for the A-configuration
 at $\omega = 0.7$\,MeV and $\theta=0^\circ$.
The Routhians calculated for the two
functionals, SkM* and SLy4$_L$, are rather similar.
Clearly visible two minima at $\gamma = -\arctan Q_{22}/Q_{20} \approx  14^{\circ}$ and $-10^\circ$ are well
separated by a potential barrier of $\sim 1$\,MeV
at $\gamma \approx 0^\circ$.  The SHFPAC Routhians
are similar to those of the SCTAC and CNS calculations of Ref.~\cite{wang11}, which have a somewhat smaller barrier of $\sim$0.7\,MeV.

\begin{figure} \centering
	\includegraphics[width=0.45\textwidth]{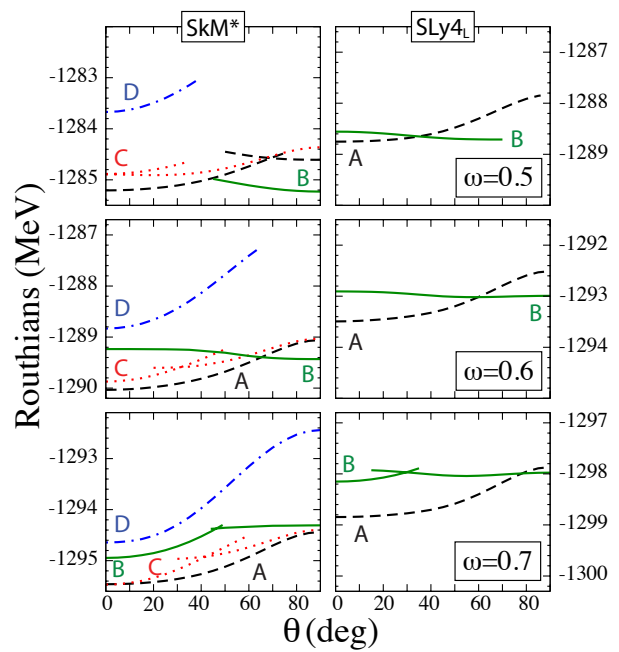}
        \caption{(Color online) Total Routhians of $^{158}$Er as a
        function of $\theta$ at $\omega$=0.5, 0.6, and 0.7\,MeV for
        SkM* (left) and SLy4$_L$ (right) functionals.
        Dashed, solid, dotted, and dot-dashed lines represent TSD configurations A, B, C, and D,
        respectively (see Table~\protect\ref{tab1}). Note that the labeling of
        the configurations is only valid when $\theta =0^{\circ}$ or
        $90^{\circ}$. Otherwise, the signature is no longer a good
        quantum number. For the SLy4$_L$ functional, converged solutions for configurations C and D
        could not be obtained.}
	\label{routhian-theta}
\end{figure}

In Fig.~\ref{routhian-theta} we allow the
rotational axis to tilt ($\theta \ne 0$)
by starting from SHFPAC solutions with
$Q_{22} \approx-4$\,b. [A rotation of this shape around the $y$-axis
($\theta=90^{\circ}$) is equivalent to that of $Q_{22} \approx 4$\,b
around the $x$-axis ($\theta=0^{\circ}$).]
It can be seen that for the configuration A, the minimum
that appears in Fig.~\ref{PES} at $Q_{22} \approx 4$\,b
($\gamma \approx - 10^{\circ}$)  is unstable
with respect to a reorientation of the rotational axis, that is, it represents the  saddle-point.
On the other hand, the lower minimum, at $\gamma \approx 14^{\circ}$, remains stable. This is consistent with SCTAC calculations of
Fig.~\ref{shape}. A similar situation is predicted for configurations C and D.
At $\omega$=0.5\,MeV,  the configuration B has a minimum at $\theta=90^\circ$
($\gamma <0$), but it becomes $\theta$-unstable  at higher rotational frequencies and a minimum at $\theta=0^\circ$ develops at $\omega$=0.7\,MeV.
This interesting behavior, together with a discussion of wobbling modes in bands A-D will be discussed in detail in a forthcoming paper.

A detailed search at larger deformations has resulted in
configuration D  having
$Q_{20} \approx 29$\,b, $Q_{22} \approx -6$\,b ($\gamma =12^\circ$).
This structure has similar occupations as configuration B, see   Table~\ref{tab1}, but much larger deformation. The calculated
$Q_t \approx 10.5$\,eb
well reproduces the experimental value of
$\sim$11\,eb~\cite{wang11}. We note that band D has a larger quadrupole moment and smaller $\gamma$ than band TSD3 of Ref.~\cite{wang11}.

\begin{figure}
\centering
\includegraphics[width=0.45\textwidth]{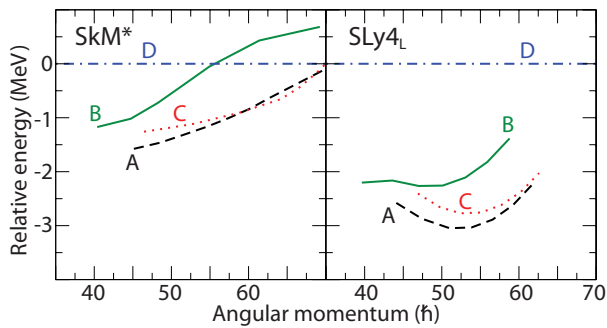}
        \caption{(Color online) Relative energies of configurations A (dashed), B (solid), and C (dotted)
        with respect to D (dash-dotted line at 0 MeV).
}
	\label{E-I}
\end{figure}
Figure~\ref{E-I} compares the
energies of configurations A, B, and C with that of D.
For the {\SKM} functional (left), which has been partly optimized at large
deformations, it can be seen that at low spins
band D lies  about 1\,MeV above band  B and 1.5-1.8\,MeV above bands A and C.
Due to its large moment of inertia, band D
crosses band B at about $J \approx 55\hbar$ and bands
A and C  at $J \approx 70 \hbar$. This is not inconsistent with the spin estimates of Ref.~\cite{paul07} based on the feeding analysis of yrast states. While we do not expect the relative  energies calculated  in SHF using current functionals to be precise, as evidenced by appreciable differences
between {\SKM} and {\SLY} predictions in Fig.~\ref{E-I}, band D is clearly
the best candidate for the  structure observed in experiment.
As seen in Fig.~\ref{routhian-theta},  D-configuration is perfectly stable against the rotational axis tilting.
\begin{table}
\caption{Charge quadrupole moments, transition (charge) quadrupole moments, and angular momenta for bands A and D calculated with {\SKM}.
}
\begin{ruledtabular}
\begin{tabular}{cccccc}
      Band &   $\omega$ (MeV) & $Q^{\text{ch}}_{20}$ (eb) & $-Q^{\text{ch}}_{22}$ (eb) & $Q_t$ (eb) &
        $J$ ($\hbar$)   \\
\hline
\multirow{4}{*}{A} &
 0.50 & 9.01 & 1.92 & 7.90 & 45.13   \\
 & 0.60 & 8.72 & 1.99 & 7.57 & 51.61 \\
 & 0.70 & 8.42 & 2.04 & 7.24 & 57.33  \\
 & 0.80 & 8.09 & 2.10 & 6.88 & 62.84 \\
 \hline
\multirow{5}{*}{D} &
 0.40 & 11.81 & 2.30 & 10.48 & 31.51 \\
 & 0.50 & 12.22 & 2.40 & 10.83 & 48.36 \\
 & 0.60 & 12.13 & 2.45 & 10.72 & 54.82 \\
 & 0.70 & 12.05 & 2.48 & 10.62 & 61.75 \\
 & 0.80 & 11.96 & 2.53 & 10.50 & 71.90 \\
\end{tabular}
\end{ruledtabular}
\label{tab}
 \end{table}
In Table~\ref{tab} we list the quadrupole moments  of the
lowest stable TSD configuration (A-configuration) at positive-$\gamma$ value
and different rotational frequencies. $Q_t$ is calculated from charge
quadrupole moments through the relation $Q_t = Q^{\text{ch}}_{20} +
\sqrt{\frac{1}{3}} Q^{\text{ch}}_{22}$ \cite{mate07}.

In summary, we have performed, for the first time, TAC calculations within the self-consistent Skyrme Hartree Fock model in which the KO conditions for triaxial rotation are strictly obeyed.  To address the recent puzzling experimental data,  we studied the nucleus $^{158}$Er at ultrahigh spins. Restricting the direction of the rotational axis to one of the principal axes of the density distribution yields two TSD minima with similar $\epsilon_2$ values but with  positive and negative $\gamma$ deformations,  similar to our SCTAC predictions  and the previous  CNS calculations~\cite{wang11}. Allowing the rotational axis to tilt away from the principal axes  shows, however,  that the higher-energy  minimum is actually a saddle point; hence, it cannot be associated with a physical state. It is the  lower-energy  minimum that represents  a  TSD  band. We have thus clarified a long-standing question pertaining to the nature of positive- and negative-$\gamma$ bands associated with the same intrinsic shape in the PAC approach: the rotation of a
well-deformed, slightly  triaxial configuration can  be either about a short or medium  axis, but not about both.

Several TSD configurations differing by proton and neutron occupations and quadrupole moments have been investigated. In the angular momentum range of 50--70 $\hbar$, they are predicted to  have transition quadrupole moments of 7--8\,eb, which are below the  measured  values of  $Q_t \approx 11$\,eb \cite{wang11}. We have identified an excited TSD configuration with a  stable positive-$\gamma$ minimum, which has   a large transition quadrupole moment of $Q_t \approx 10.5$\,eb that agrees well with the experimental value. At spins higher than $\sim 55\hbar$, this band -- different from structures TSD3 and SD of Ref.~\cite{wang11} -- lies close to the less deformed TSD bands, and it is expected to become yrast above $J> 70 \hbar$.

Pertinent and stimulating questions by Mark Riley, and numerous valuable discussions with him, are gratefully acknowledged.
This work has been supported by the Natural Science
Foundation of China under Grants Nos. 10735010 and 10975006;
U.S. Department of Energy under
Contract Nos.  DE-FG02-96ER40963
 (University of Tennessee) and  DE-FG02-95ER40934 (University of Notre Dame);
Academy of Finland and the
University of Jyv\"askyl\"a within the FIDIPRO programme.


\end{document}